\documentclass[prb,twocolumn,showpacs,superscriptaddress]{revtex4}
\usepackage[dvips]{graphicx}
\usepackage{dcolumn}
\usepackage{bm}
\usepackage{amssymb}

\DeclareGraphicsExtensions{eps}

\begin{document}
\preprint{APS/123-QED}
\title{Low temperature magnetization of the S=1/2 kagome antiferromagnet ZnCu$_{3}$(OH)$_ {6}$Cl$_{2}$}
\author{F. Bert}
\affiliation{%
Laboratoire de Physique des Solides, UMR CNRS 8502, Universit\'{e}
Paris-Sud, 91405 Orsay, France}%
\author{S. Nakamae}
\affiliation{Service de Physique de l'\'Etat Condens\'e, DSM, CEA
Saclay, 91191 Gif-sur-Yvette Cedex, France.}
\author{F. Ladieu}
\affiliation{Service de Physique de l'\'Etat Condens\'e, DSM, CEA
Saclay, 91191 Gif-sur-Yvette Cedex, France.}
\author{D. L'H\^ote}
\affiliation{Service de Physique de l'\'Etat Condens\'e, DSM, CEA
Saclay,
91191 Gif-sur-Yvette Cedex, France.}
\author{P. Bonville}
\affiliation{Service de Physique de l'\'Etat Condens\'e, DSM, CEA
Saclay, 91191 Gif-sur-Yvette Cedex, France.}
\author{F. Duc}
\affiliation{%
Centre d'\'Elaboration des Mat\'eriaux et d'\'Etudes Structurales,
CNRS UPR 8011, 31055 Toulouse,
France}%
\author{J.-C. Trombe}
\affiliation{%
Centre d'\'Elaboration des Mat\'eriaux et d'\'Etudes Structurales,
CNRS UPR 8011, 31055 Toulouse,
France}%
\author{P. Mendels}
\affiliation{%
Laboratoire de Physique des Solides, UMR CNRS 8502, Universit\'{e}
Paris-Sud, 91405 Orsay, France}%

\date{\today}

\begin{abstract}
The dc-magnetization of the unique S=1/2 kagome antiferromagnet
Herbertsmithite has been measured down to 0.1~K. No sign of spin
freezing is observed in agreement with former $\mu SR$ and
ac-susceptibility results. The low temperature magnetic response is
dominated by a defect contribution which exhibits a new energy scale
$\simeq 1$~K, likely reflecting the coupling of the defects. The
defect component is saturated at low temperature by $H\gtrsim 8$~T
applied magnetic fields which enables us to estimate an upper bound
for the non saturated intrinsic kagome susceptibility at $T=1.7$~K.
\end{abstract}

\maketitle

In triangular lattices, the frustration of antiferromagnetic
interactions associated to the enhancement of quantum fluctuations
for S=1/2 spins was acknowledged long ago as a keypoint to stabilize
novel ground states of magnetic matter~\cite{Anderson73}. Numerous
theoretical studies have since then emphasized the S=1/2
nearest-neighbor Heisenberg antiferromagnet on the kagome lattice
(KAH), a network of corner sharing triangles. Although numerical
approaches are complicated by the huge degeneracy of the system, it
is believed that the ground state could be a unique realization of a
disordered two dimensional quantum liquid at $T=0$, with a
surprisingly small gap, if any, to unconventional unconfined spinon
excitations and a gapless continuum of non magnetic
excitations~\cite{Lecheminant97,Waldtmann98,Misguich03}.
Concurrently, growing efforts were made to identify a model
frustrated compound and find evidences for such an exotic spin
liquid ground state. Key features have emerged from these
experimental investigations like the suppression of magnetic order
at the energy scale of the antiferromagnetic interaction, the
persistence of spin dynamics at very low
temperatures~\cite{Uemura94,Bono04b} or a large density of low
energy non magnetic states~\cite{Ramirez00}. However, the frustrated
compounds studied so far show strong deviations from the ideal KAH
(S$>1/2$ spins, dilution of the magnetic lattice, anisotropic
interactions). Besides, they often present marginal low $T$ order or
spin glass like behavior which forbid a close comparison to
theoretical expectations~\cite{Bert05}.

Only very recently, Herbersmithite, ZnCu$_{3}$(OH)$_6$Cl$_2$, a
structurally perfect kagome antiferromagnet decorated by Cu$^{2+}$
S=1/2 spins could be synthesized~\cite{Shores05}. It belongs to a
large compound family Zn$_x$Cu$_{4-x}$(OH)$_6$Cl$_2$ where the
parent structure, clinoatacamite (x=0), is that of a distorted S=1/2
pyrochlore. The substitution of Zn$^{2+}$ ions preferentially on the
less Jahn-Teller distorted Cu$^{2+}$ site located in between the
kagome planes restores the three fold symmetry of the lattice for
$x>1/3$.  Eventually the Herbertsmithite compound ($x=1$) presents
decoupled S=1/2 perfect kagome planes. Also, the magnetic order
which sets in clinoatacamite at 19~K gradually disappears as $x
\rightarrow 1$ and for $x=1$, muon spin resonance ($\mu$SR)
investigation~\cite{Mendels07} has demonstrated the absence of any
spin freezing at least down to 50~mK, an energy scale 4000 times
smaller than the main antiferromagnetic interaction ($J\simeq
190$~K).

Once Herbertsmithite is acknowledged to be the first good candidate
for the realization of the KAH model, the magnetic susceptibility
and heat capacity are the first quantities of interest as they
straightforwardly probe the nature of the ground state, either
magnetic or not, and the excitation spectrum. At low $T$, these
thermodynamic quantities show respectively a Curie-like
tail~\cite{Helton07,Ofer07} and a
Schottky-type~\cite{Helton07,deVries07} anomaly. It was soon
recognized that Dzyaloshinsky-Moriya interactions, which are allowed
in Herbertsmithite structure, can yield such a drastic increase of
the low $T$ susceptibility~\cite{Rigol07}. However magnetic defects
could also account for these features, a scenario sustained by
recent NMR data~\cite{Imai07} and neutron diffraction refinements of
the structure~\cite{deVries07,Lee07}. Despite a poor sensitivity,
these latter point at a large ($6-10\%$) Cu/Zn intersite mixing.
This chemical disorder would likely reflect the finite energy of the
Jahn-Teller process that selects the Zn substitution site.
 Both the resulting dilution of the kagome magnetic network and
the interplane Cu$^{2+}$ ions may contribute to the defect
component.  In this paper, we report on a detailed investigation of
the Herbertsmithite magnetization at low temperature ($T > 0.1$~K)
and up to moderately high fields ($H < 14$~T). Both sets of data are
consistently analyzed in terms of a large defect contribution. We
show that the low $T$ intrinsic susceptibility can be nonetheless
estimated.

A ZnCu$_{3}$(OH)$_ {6}$Cl$_{2}$ powder sample was prepared by the
hydrothermal method described in Ref.~\cite{Shores05,Mendels07}. Low
temperature dc-magnetization (0.1~K$<T<3$~K) was measured in a home
made SQUID magnetometer for fixed external magnetic fields up to
0.7~T. For $T>0.3$~K, each data point was obtained by extracting the
sample through the pick-up coils. To avoid heating effect, for
0.1~K$<T<0.4$~K, the sample was kept at a fixed position in the
pick-up coils and we measured the SQUID voltage variation as a
result of the $T$-dependent sample magnetization. Besides,
dc-magnetization curves were recordered versus field ($0<H<14$~T) at
constant temperature in a commercial vibrating sample magnetometer
(VSM). Standard SQUID data up to 5~T and for $T>$1.8~K were also
used to complement and calibrate the low $T$ data.

\begin{figure}[!t]
\includegraphics[scale=0.8]{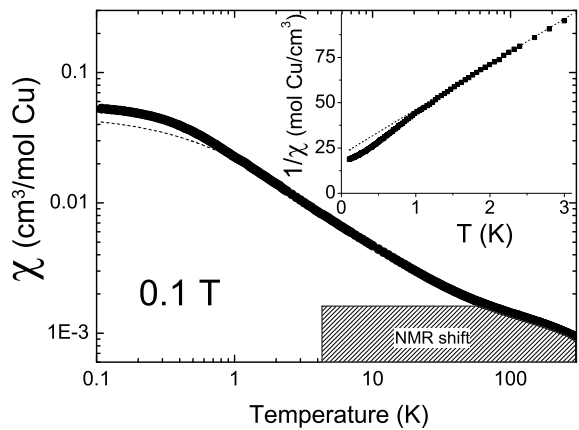}
\caption{\label{fig1} Molar dc-susceptibility of Herbertsmithite
versus temperature in 0.1~T external field on a log-log plot. The
local susceptibility measured by $^{35}$Cl NMR line shift falls in
the shaded area~\cite{Imai07}. The dashed line is a Curie Weiss fit
for 1.5~K$<T<3$~K (see text). Inset: $T$-dependence of the inverse
of the susceptibility at low temperature. }.
\end{figure}

\begin{figure}[!t]
\includegraphics[scale=0.8]{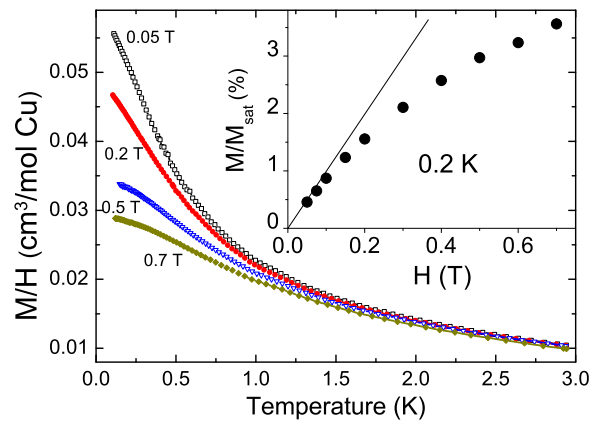}
\caption{\label{Fig2} (Color online) $M/H$ as a function of
temperature for different $H$. Inset: $M$ normalized by the
saturated magnetization of one mole of S=1/2 spins (Cu$^{2+}$)
$M_{sat}=5583$~emu, at 0.2~K as a function of $H$. The data are
extracted from the temperature scans of the main panel. The straight
line is a plot of $\chi(0.2$~K$)H$.}
\end{figure}

The temperature dependence of the dc magnetic susceptibility $\chi$
of Herbertsmithite measured in a 0.1~T applied field is presented on
Fig.~\ref{fig1} in the whole studied temperature range
(0.1~K$<T<300$~K). At high temperature $T\gtrsim$150~K), the
susceptibility shows a Curie-Weiss behavior which yields the
exchange constant $J\simeq 190$~K\cite{Helton07,Misguich07}. At
lower temperature, the susceptibility increases much more rapidly
down to $\simeq 0.5$~K where it eventually flattens. Down to the
lowest temperature of the experiment $T=0.1$~K, there is no sign of
a magnetic transition in agreement with former
$\mu$SR~\cite{Mendels07} and ac-susceptibility~\cite{Helton07}
measurements. The $T$-dependence of the total dc-magnetization $M$
has been also measured between 0.1~K and 3~K for fixed applied
fields $H$ in the range 0.05 - 0.7~T. Characteristic plots of $M/H$
versus $T$ are presented in Fig.~\ref{Fig2}. At low temperature,
saturation effects are evidenced by the decrease of $M/H$ with
increasing fields. More precisely, the field dependence of the
magnetization measured at 0.2~K is plotted in the inset of
fig.~\ref{Fig2}. At this temperature, the data are well described by
the linear $M=\chi(0.2$~K$)H$ relation for low fields $H\leq 0.1$~T
while saturation effects are clearly observed for $H \geq 0.2$~T.
Therefore, the flattening at low temperature of $M/H \simeq \chi$
measured for $H=0.1$~T in Fig.\ref{fig1} can not be ascribed to a
field effect.

The NMR lineshift of chlorine in Herbertsmithite was measured
recently~\cite{Imai07}. This local probe investigation is thought to
give the intrinsic susceptibility $\chi_i$ which is found to
strongly deviate from the macroscopic SQUID data $\chi$ for $T
\lesssim 50$~K. Indeed, despite large error bars due to the
broadening of the NMR line, $\chi_i$ shows a broad maximum or at
least a saturation below $\simeq 50$~K. One can then put an upper
limit to the intrinsic susceptibility $\chi_i < 1.5 \times
10^{-3}$~cm$^3$/mol Cu as represented by the shaded area in
Fig.~\ref{fig1}. The maximum of $\chi_i$ likely reflects a moderate
enhancement of the short range AF correlations as in the well
studied kagome bilayer case~\cite{Mendels00,Mondelli99,Bono04}. Once
these correlations have developed, it is doubtful that there will be
a subsequent rise of the susceptibility at lower temperature and we
assume that the above upper limit for $\chi_i$ also stands down to
0~K. The low temperature dc-susceptibility which is the subject of
this report is therefore mainly dominated by a defect contribution.
In the following we will make the simplest assumption that the
intrinsic and defect contributions are uncorrelated and therefore
$\chi=\chi_d+\chi_i$ with $\chi_i/ \chi_d < 0.1$ for $T<2$~K. As
previously mentioned, one can anticipate two types of defects in the
structure which can both show a paramagnetic-like behavior and which
both contribute to $\chi_d$ in our analysis. First some Cu$^{2+}$
ions could lie on the interplane site. Their coupling to the kagome
planes is likely very weak, maybe slightly ferromagnetic, as
discussed in Ref.~\cite{Shores05}. Second, the dilution of the
kagome magnetic lattice by Zn$^{2+}$ ions is believed to locally
stabilize dimers and to induce a weak staggered magnetization on
further neighboring sites~\cite{Dommange03}. This non trivial
extended response of the system around a spin vacancy constitutes
the second magnetic defect. In the closely related copper based
anisotropic kagome structure of Volborthite~\cite{Hiroi01}, the
controlled magnetic dilution by Zn/Cu substitution indeed yields a
Curie-like tail that scales with the Zn content~\cite{Bert04}.

We first consider the intermediate temperature range 1.5~K~--10~K.
 As shown in the inset of Fig.\ref{fig1}, $1/\chi \simeq 1/\chi_d$ does not extrapolate to 0
when $T\rightarrow 0$ as would be expected for free spins following
a Curie law. Instead $\chi_d$ rather shows a Curie-Weiss behavior
$\chi_d=C_d/(T+\theta_d)$. A proper fit of the low T data requires
an accurate knowledge of $\chi_i (T)$. In the absence of such data,
we fit with a constant $\chi_i$ in the two extreme cases; $\chi_i=0$
which yields $\theta_d=0.85$~K and $C_d =0.040$~cm$^3$/mol Cu/K (fit
range 1.5~K$<T<3$~K) and $\chi_i=1.5 \times 10^{-3}$~cm$^3$/mol Cu
which yields $\theta_d=0.80$~K and $C_d =0.0345$~cm$^3$/mol Cu/K
(fit range 2~K$<T<10$~K). A fit of the high temperature ($T>150$~K)
data gives a Curie-Weiss constant $C_{CW}\simeq 0.5$~cm$^3$/mol Cu.
If one assumes that the magnetic defects behave as S=1/2 spins,
their contribution corresponds to $\backsim 7\%$ of weakly coupled
S=1/2 spins out of the total Cu$^{2+}$ contribution. This number is
remarkably similar to the estimated number of two level systems
which contribute to the Schottky anomaly in heat capacity
measurements~\cite{deVries07} and also of misplaced Cu$^{2+}$ from
neutron diffraction refinement~\cite{deVries07,Lee07}. This latter
finding suggests that the main contribution to $\chi_d$ comes from
the interplane Cu$^{2+}$ (S=1/2 defects) whereas the integrated
staggered magnetization around a Zn$^{2+}$ amounts to a rather small
moment.

More puzzling is the behavior below 1~K where a subsequent
enhancement of the susceptibility appears (Fig.~\ref{fig1}, inset).
The above described Curie-Weiss regime accounts then only
qualitatively for the flattening of $\chi(T)$ (see dashed line in
main panel and inset). At 0.1~K the rise of $\chi$ with respect to
the extrapolated Curie-Weiss behavior is about $1.3 \times
10^{-2}$~cm$^3$/mol Cu, \emph{i.e.} one order of magnitude larger
than the upper limit of $\chi_i$.  This enhancement is therefore
also related to the defect contribution $\chi_d$. No Field
Cooling-Zero Field Cooling opening could be detected below 1~K.
Moreover $\chi$ does not show any peak or divergence that would
signal long range ordering. Thus, the rise of $\chi$ for $T \simeq
\theta_d$, probably reflects a strengthening of some
ferromagnetic-like correlations between the magnetic defects rather
than
 some kind of ordering. It is noticeable that a slight slowing down
 of the electronic spin fluctuation is detected at this same
 temperature $T \simeq \theta_d$ in $\mu$SR experiments. This nicely corroborates the
 correlation strengthening picture.  $\theta_d$ is also
 close to the temperature of the maximum of the Schottky anomaly in
 zero field heat capacity data. Therefore, $k_B \theta_d$ appears as
 a new energy scale for Herbertsmithite, most likely related to the
 magnetic defect system.

\begin{figure}
\includegraphics[scale=0.8]{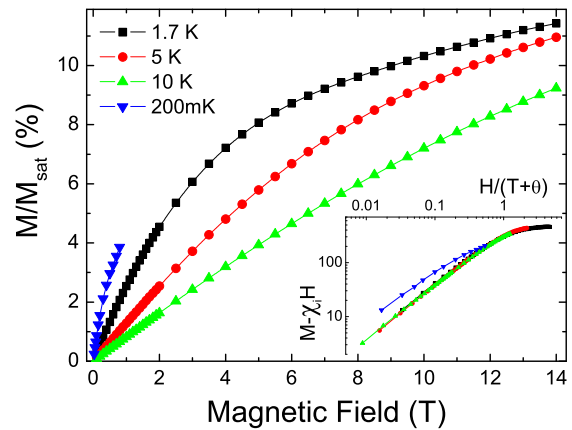}
\caption{\label{fig3} (Color online) Normalized magnetization
($M_{sat}=5583$~emu) measured versus field in a VSM for 3
characteristic temperatures. The magnetization data at 0.2~K are
also reported from the inset of Fig.~\ref{Fig2}. In the inset, for
the same temperatures, $M-\chi_iH$ versus $H/(T+\theta)$ with
$\chi_i=1.25 \times 10^{-3}$~cm$^3$/mol Cu and $\theta=1.3$~K.}
\end{figure}

\begin{figure}
\includegraphics[scale=0.8]{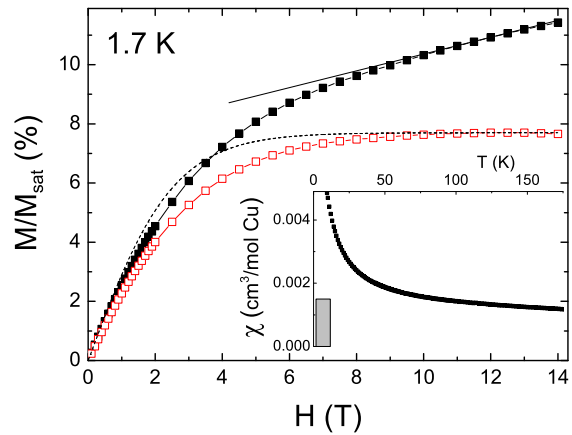}
\caption{\label{fig4} (Color online) Full squares : normalized
magnetization of Herbertsmithite measured at 1.7~K versus field. The
solid line is a linear fit of the data for $H>10$~T which likely
reflects the non saturated intrinsic susceptibility $\chi_i$. Open
squares : the defect contribution obtained by subtracting the above
linear contribution from the full square magnetization curve. Dashed
line : Brillouin function for 7.7\% of free S=1/2 spins. Inset: the
shaded area on the SQUID $\chi(T)$ plot represents the possible
values of $\chi_i$ from this study.}
\end{figure}

The magnetic response of the defects strongly dominates the total
susceptibility at low temperature and it is difficult to extract any
information on the KAH contribution. However, one can expect
different field dependences for the two contributions. Namely, the
weakly coupled magnetic defects should be more easily saturated than
the Cu$^{2+ }$ spins belonging to the perfect kagome network with
the stronger $J \simeq 190$~K coupling. To further investigate the
field dependence of the magnetization we measured $M(H)$ curves in a
vibrating sample magnetometer up to 14~T for constant temperatures
in the range 1.7~K--25~K. Characteristic results are shown in
Fig.~\ref{fig3}. At 1.7~K, a strong saturation effect is observed
above $\sim$2~T and up to $\sim$8~T where $M(H)$ reaches a linear
regime. At higher temperature, the saturation effect gradually
disappears and at 10~K a nearly linear dependence is recovered. This
behavior is compatible with a simple decomposition of the
magnetization into a defect and an intrinsic contribution
$M=M_d+M_i$. One then assumes that for moderately high fields
($H<14$~T) with respect to the coupling energy scale $J$, the linear
$M_i(H,T)=\chi_i(T)H$ relation holds and that the defect
magnetization $M_d$ follows a Brillouin like saturation. In this
scenario, at 1.7~K, the regime for $H>10$~T where the magnetization
is a linear function of $H$, is explained by the complete saturation
of the magnetic defects and the slope of $M(H)$ is a direct measure
of the intrinsic susceptibility $\chi_i(1.7$~K). In Fig.~\ref{fig4}
the straight line corresponds to $\chi_i(1.7$~K$)=1.5 \times
10^{-3}$~cm$^{3}$/mol Cu. It is the largest possible value for the
intrinsic susceptibility so that the remaining defect magnetization
extracted from our data (open squares) does not decrease at high
fields. It is noticeable that the fully saturated defect
magnetization amounts then to $\sim 8\%$ of the saturated
magnetization of one Cu$^{2+}$ mole, in perfect agreement with the
$\sim 7\%$ estimate given by the low T Curie-Weiss behavior of the
defect susceptibility. However, one cannot exclude that the complex
magnetic defects at play in Herbertsmithite are not completely
saturated even at the lowest temperature and highest field of this
study. Part of or the whole linear regime could then be ascribed to
the defect contribution. The extracted $\chi_i$ value is therefore
only the upper limit of the kagome susceptibility at low
temperature. The possible values of $\chi_i$ for $T\gtrsim 1.7$~K
from this analysis are represented by the shaded area in the inset
of Fig.~\ref{fig4}.

As shown by the dashed line in Fig.~\ref{fig4}, a simple S=1/2
Brillouin function fails to capture the field dependence of the
defect magnetization $M_d=M-\chi_iH$. Possible reasons for this are
that 1) the magnetic defects are complex objects involving the point
defect itself, likely a misplaced Zn/Cu atom, and the local
screening of the defect by the neighboring spins, so that one does
not expect a simple S=1/2 effective spin value, 2) the magnetic
defects are slightly antiferromagnetically coupled which tends to
reduce the field effect with respect to the $H/T$ dependence of free
spins. Note that unconstraining the spin value of the Brillouin
function does not give either a good fit of the $H$ and $T$
dependence of $M_d$. As shown in the inset of Fig.~\ref{fig3}, the
$M_d(T,H)$ data for $1.7$~K$<T<10$~K merge on a same curve if one
uses the scaling variable $H/(T+\theta)$ which accounts
phenomenologically for the AF coupling. Good scaling is obtained for
$\chi_i=0.00125 \pm 0.00025$~cm$^{3}$/mol Cu and $\theta=1.1 \pm
0.2$~K in agreement with $\theta_{CW}$ extracted from the low T
Curie Weiss fit of the susceptibility. It is noticeable that below
1~K, deviations from this scaling appear gradually. Eventually, the
0.2~K curve can not be made to fall on the $T>1$~K ones, even with
different $\chi_i$ and $\theta$ values. It suggests that the
effective defect moment does change below 1~K which corroborates the
enhanced correlations scenario drawn from the analysis of the $T$
dependence of the susceptibility.

In summary, from a detailed study of the temperature and field
dependence of the magnetization at low $T$ , we can draw a coherent
picture of the Herbertsmithite magnetic behavior. The Curie-like
tail in the susceptibility can be safely attributed to a defect
contribution. The magnetic defects, probably of two kinds, behave in
average as weakly coupled spins ($S\neq1/2$). Signature of the
coupling energy $\simeq 1$~K are found ubiquitously in
thermodynamics measurements as well as in the spin dynamics. The
complex nature of the defects challenges both chemistry to achieve a
better control of Zn/Cu site occupation and theory to describe their
magnetic behavior. Remarkably, such a large quantity of defects does
not seem to alter the underlying KAH physics. Besides, although the
effect of the $\simeq 10$~T external fields used in this study is
not clearly known, our results are compatible with a finite kagome
susceptibility at $T \simeq J/100$ and thus question the ground
state nature and the presence of a gap. Although we showed in this
study that the low $T$ up turn of the macroscopic susceptibility can
be explained in a defect scenario without Dzyaloshinsky-Moriya
perturbation terms contrary to the initial proposal of
Ref.~\cite{Rigol07}, they could nonetheless impact the low $T$
intrinsic properties and possibly increase the polarisability of the
ground state of this unique realization of a S=1/2 kagome system.


\end{document}